\documentstyle[12pt]{article}
\newcommand{\be}{\begin{eqnarray}}
\newcommand{\ee}{\end{eqnarray}}

\newcommand{\ra}{\rightarrow}
\begin{document}
\hfill IFT/95/6

\hfill June 1995

\hfill hep-ph/9506291

\begin{center}
{\large \bf Fermion mass effects on ${ \Gamma(Z\ra b\bar{b} + a~light~Higgs)}$
\\
in a two-Higgs-doublet model
  } \\

\vspace{1cm}
J. Kalinowski\footnote{Supported in part by the grant 2 P302 095 05
from Polish Committee for Scientific Research}, M. Krawczyk \\
Insitute of Theoretical Physics \\
Warsaw University \\ ul. Ho\.{z}a 69,
00 681 Warsaw,
Poland
\end{center}
\vspace{1cm}
\begin{abstract}
Large fermion mass effect on the Yukawa process $Z \ra b\bar{b}\ra b\bar{b} h
(A)$
 in the entire range of the neutral Higgs boson masses is found. It is
particularly
important for light Higgs bosons, which are still not excluded experimentally
in a general
two-Higgs-doublet model.
\end{abstract}
\vspace{1cm}

The light Higgs bosons in the context of the standard model (SM)
and its minimal supersymmetric extension (MSSM) have been
ruled out at LEP from zero up to a value that depends on
the model assumed. In the framework of the SM the dominant Higgs boson
production
process is the Bjorken process
\be
Z \ra H_{SM}Z^* \ra H_{SM}f\bar{f} \label{bj}
\ee
and the mass limit $m_{H_{SM}}\geq 65.1$ GeV has been established based on the
results of 4 collaborations collecting data at LEP\cite{4coll}.
In extensions of the SM, like MSSM or general two-Higgs-doublet model (2HDM),
the Higgs sector is much richer: there are two CP-even neutral
Higgs bosons denoted
by $h$ and $H$ (we assume that $m_h\leq m_H$), a CP-odd neutral -- $A$, and
a pair of charged scalars $H^{\pm}$.
The CP-even Higgs bosons can be produced via the process (\ref{bj})
however with lower rates because they share the couplings to $Z$ boson,
$i.e.$
\be
\Gamma(Z\ra hZ^*)&=&\Gamma_{SM}(Z\ra H_{SM}Z^*) \sin^2(\beta-\alpha)
\label{bjh}\\
\Gamma(Z\ra HZ^*)&=&\Gamma_{SM}(Z\ra H_{SM}Z^*) \cos^2(\beta-\alpha)
\label{bjH}\ee
where $\alpha$ and $\beta$ are mixing angles in the neutral and charged
Higgs sectors, respectively, with $\tan\beta$  given by the
ratio of the vaccuum expectation values of the Higgs doublets,
$\tan\beta=v_2/v_1$.

In both MSSM and 2HDM models there is another production process,
namely the Higgs pair production, which is complementary to (\ref{bjh}) and
(\ref{bjH})
in the sense that
\be
\Gamma(Z\ra hA)=0.5 \Gamma(Z\ra\nu\bar{\nu}) \cos^2(\beta-\alpha)
\lambda^p \label{hph}\\
\Gamma(Z\ra HA)=0.5 \Gamma(Z\ra\nu\bar{\nu}) \sin^2(\beta-\alpha)
\lambda^p \label{hpH}
\ee
where $\lambda=(1-\kappa_h-\kappa_A)^2-4\kappa_h\kappa_A$,  with
$\kappa_i=m_i^2/m_Z^2$ and $p=3/2$.

In the MSSM the parameters of the Higgs sector are constrained and only
two of them are independent, for example $\tan\beta$ and $m_A$.
This remains true also after
taking radiative corrections, which are important for a heavy top quark,
although the relations among them are modified and  depend
on $m_t$ and some other parameters of the supersymmetric sector.
The allowed domain for
$m_h$ and $m_A$ is restricted theoretically and
for any given $m_h$ and $m_A$  the values of $\sin^2(\beta-\alpha)$ and
$\cos^2(\beta-\alpha)$ are restricted to vary in a certain
range.
In the MSSM the heavier Higgs boson $H$ cannot be produced at LEP
(because $m_{H}\geq m_Z$)  and  combining the negative search results
from processes (\ref{bjh}) and (\ref{hph}) the Delphi
Collaboration\cite{delphi}
recently set the mass limits $M_h\geq 44$ GeV for any $\tan\beta$ and
$M_A\geq 27$ GeV for $\tan\beta \geq 1$ assuming  the mass
of the top quark $m_t=170$ GeV and degeneracy of the top-squarks with
$m_{sq}=1$ TeV. There is, however, no lower limit on $M_A$ when
$M_h\geq 60$ GeV.

On the other hand in a general 2HDM the masses and mixing angles in the Higgs
sector
are unrelated and unconstrained theoretically and
therefore much weaker bounds can be established. From the  process (\ref{bjh})
one can derive
experimentally an upper limit  $\sin^2(\beta-\alpha) <0.1$ for $M_h<50$
GeV\cite{4coll}.
Therefore
one can imagine a scenario in which $\sin^2(\beta-\alpha) \sim 0$. In addition,
if one assumes
the
process (\ref{hph}) to be  forbidden kinematically ($i.e.$ $M_A+M_h > M_Z$),
then either very
light $h$ or very light $A$ cannot be ruled out on the basis of negative
searches via processes (\ref{bjh}) and (\ref{hph}).

For large $\tan\beta$, however, there is still
another important process in the 2HDM,\footnote{In the MSSM the process
(\ref{br})
is never competitive
with the sum (\ref{bjh})+(\ref{hph}) for $M_h\leq 50$ GeV.}
namely the bremsstrahlung of the Higgs boson from
the fermion line in the final
state,  which we will call a Yukawa process (Fig.1.)
\be Z\ra f\bar{f} \ra f\bar{f} h \;\; \mbox{or} \;\; f\bar{f} A \label{br}
\ee
 It can produce a substantial number of Higgs bosons for $f=b,\tau$
 because of strong enhancement of the Higgs couplings to down-type fermions
by a factor $\sim \tan\beta$.\footnote{For example, for $M_h=10$ GeV,
$\tan\beta=20$ one expects about 3000 $b\bar{b}h$ events in $10^7$ $Z$ decays.}
The Higgs boson then would predominantly
decay to the heaviest  possible fermion pair. Note that it is a single Higgs
boson
production process with different topology than in the Bjorken process.
Such events have not been fully analysed experimentally yet, although
the possible importance of such a process has been pointed out in the
literature long time ago\cite{sg,dzz,kp,kn}.

The process (\ref{br}) has been analysed in the literature
analytically\cite{dzz} for a pseudoscalar and numerically\cite{kp,kn} for
a scalar Higgs bosons. The analytical
treatment can however be done in the limit of vanishing fermion masses
in which the the formulas for $h$ and $A$ production are the same (after
appropriate change of the Higgs couplings and masses).

In this short note we would like point out that the proper treatment
of fermion  masses is quite important not only at the edge
of the available phase space when $2m_f+m_{h,A} \ra m_Z$ but in the entire
range of the Higgs boson mass.

The analytical expressions in the lowest order for the differential decay
distributions $\mbox{d}\Gamma/\mbox{d}x_1\mbox{d}x_2$ (with $x_1=2 E_b/M_Z$
and $x_2=2 E_{\bar{b}}/M_Z$)
for both $Z\ra b\bar{b}h$ and $Z\ra b\bar{b}A$ can be inferred from eqs.\
(15) and (17) in ref.\cite{dkz}.
In the case of $\alpha=\beta$ (which we consider here) both distributions
scale as $m_b^2 \, \tan^2\beta$
due to the Higgs coupling $\sim m_b\, \tan\beta$ to the down-type fermion line.
In addition, the fermion mass $m_b$ enters $(i)$ the phase space integration
limits
when the contribution to the total decay rate is
calculated, and $(ii)$ the matrix element. If one neglects $m_b$ in  the
matrix elements the distributions for
$Z\ra b\bar{b}h$ and $Z\ra b\bar{b}A$ are equal, and neglecting in addition
$m_b$ in $(i)$
the integral over $x_1$ and $x_2$ can be done in an elegant way\cite{dzz}.
The resulting decay widths\footnote{$\Gamma=1$ MeV corresponds to 4000 events
per $10^7$ $Z$ decays} as a function of the corresponding Higgs masses are
shown in Figs. 2 and 3 (dotted lines).
However, we find that although $m_b/M_Z \ll 1$ the fermion mass $m_b$  plays an
important
role in the entire range of Higgs boson mass.
In Figs. 2 and 3 we show the effect of retaining the fermion mass (with $m_b=5$
GeV)
in the integration limits
only and neglecting it in the matrix element (dashed lines), and keeping full
$m_b$ dependence (solid lines).

In the case of $Z\ra b\bar{b}A$ (Fig.2) we see that
keeping $m_b\not= 0$ is very important -- it cures the fake IR behaviour
of the massless approximation and results in lowering the predicted decay rate
by at least 15\% for all $M_A$. Comparing the dashed and solid lines in Fig.2
we
notice that once the kinematical limits are properly taken into account in
$(i)$
one can neglect $m_b$ in the matrix element $(ii)$.

On the other hand for $Z\ra b\bar{b}h$ (Fig.3) keeping $m_b\not= 0$ in $(i)$
and
$(ii)$ leads to an enhancement of the scalar $h$ production in $Z$ decays up to
$m_b=50$ GeV (solid line). Also, contrary to the pseudoscalar case, one cannot
neglect $m_b$ in the matrix element because it would lead to wrong IR
behaviour (dashed line).

Similar mass effects also appear
in the process (\ref{br}) with $f=\tau$ lepton in the final state.
The results  presented here have been obtained for $\tan\beta=20$.
For other values of $\tan\beta$
the results can simply be obtained by rescaling.

The observed fermion mass effects are  phenomenologically important as they
modify significantly the predicted decay rate of the $Z$ boson
as compared to the massless approximation. They have to be taken
into account when performing the detailed search for a light Higgs boson or
in deriving the experimental limits on the parameters of the
two-Higgs-doublet model\cite{kkss}.

\vspace{1cm}
\noindent {\large\bf Acknowledgments}\\
We would like to thank A. Djouadi and P. Zerwas for useful discussions.
J.K. thanks P.~Chappetta and CPT Marseille  for kind hospitality at CPT, where
part of this work has been done. M.K. is grateful to W.~Hollik and colleagues
from
Ho\.za for helpful discussions.

\vspace{1cm}
\noindent {\large \bf Figure Captions}
\begin{itemize}
\item[Fig.1.] The Feynman diagrams for the neutral Higgs boson production ($h$
or $A$)
via the Yukawa process.
\item[Fig.2.] The decay width $\Gamma(Z\ra b\bar{b}A)$ as a function of the
pseudoscalar Higgs boson mass
in the 2HDM for $\tan\beta=20$. The solid line represents the results with full
$m_b$ dependence,
dashed lines - with $m_b$ neglected in the matrix element, and dotted lines -
with $m_b$
neglected in the matrix element and integration limits.
\item[Fig.3.] The decay width $\Gamma(Z\ra b\bar{b}h)$ as a function of the
scalar Higgs boson mass
in the 2HDM for $\tan\beta=20$. Tla labeling of lines is the same as in Fig.3.
\end{itemize}
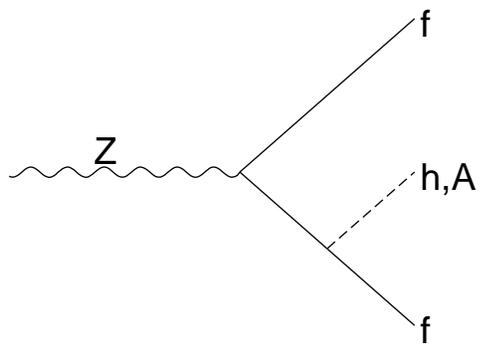
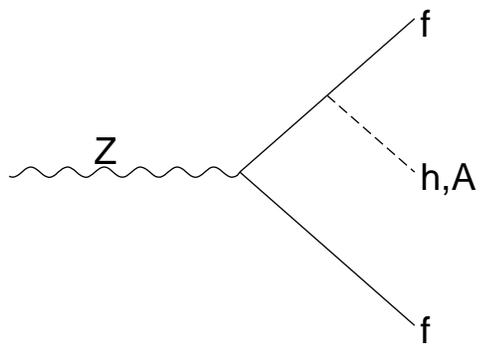
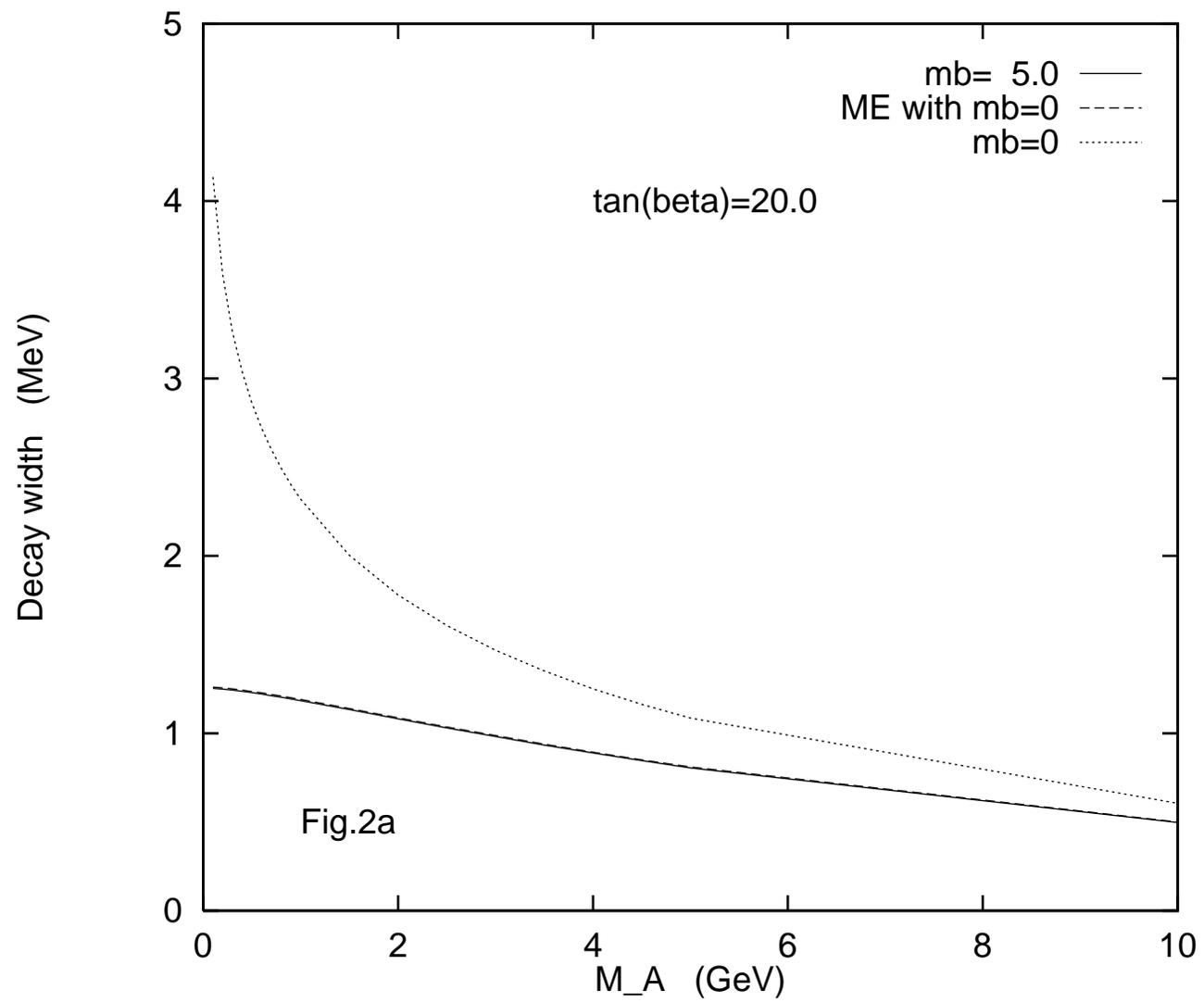
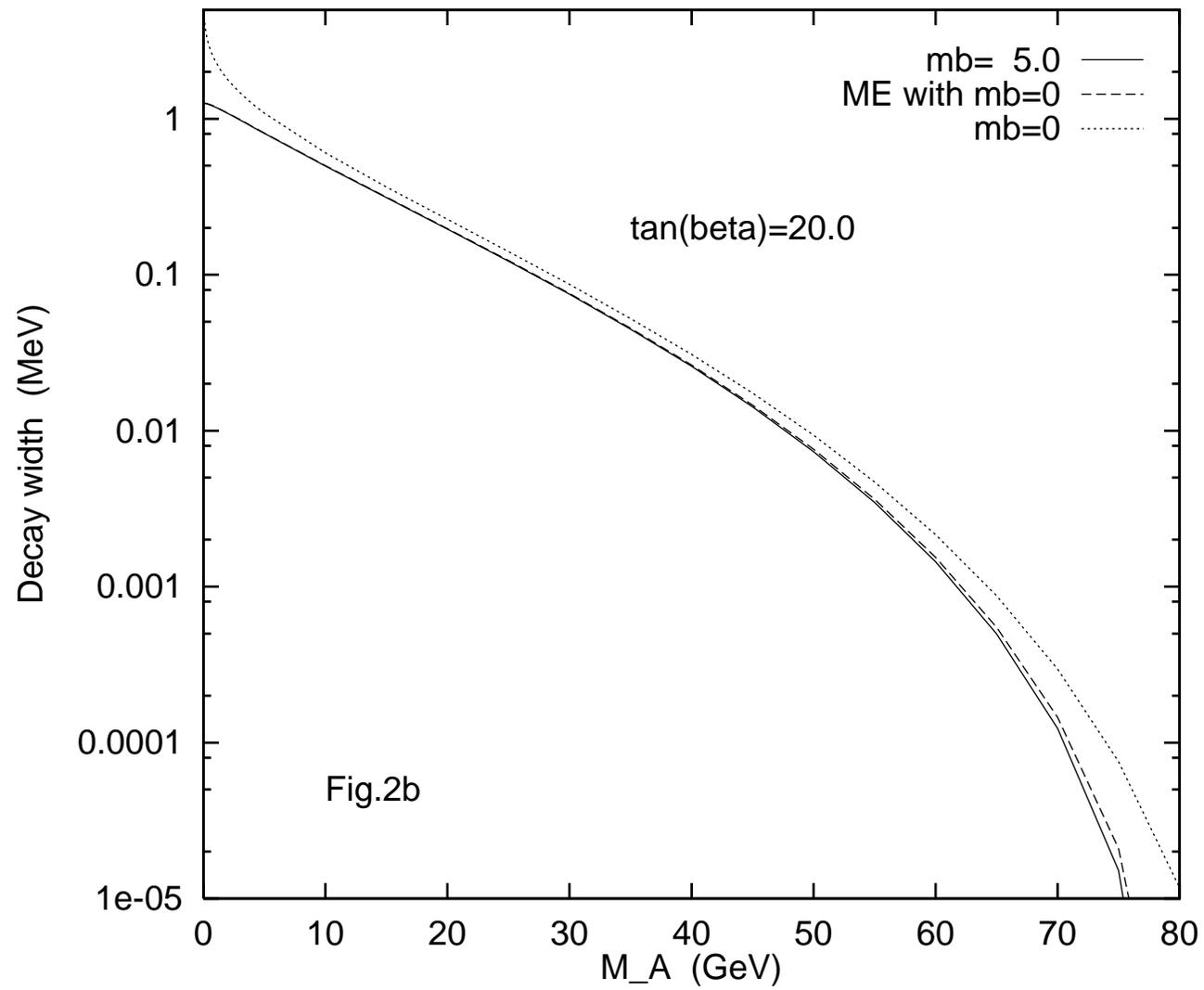
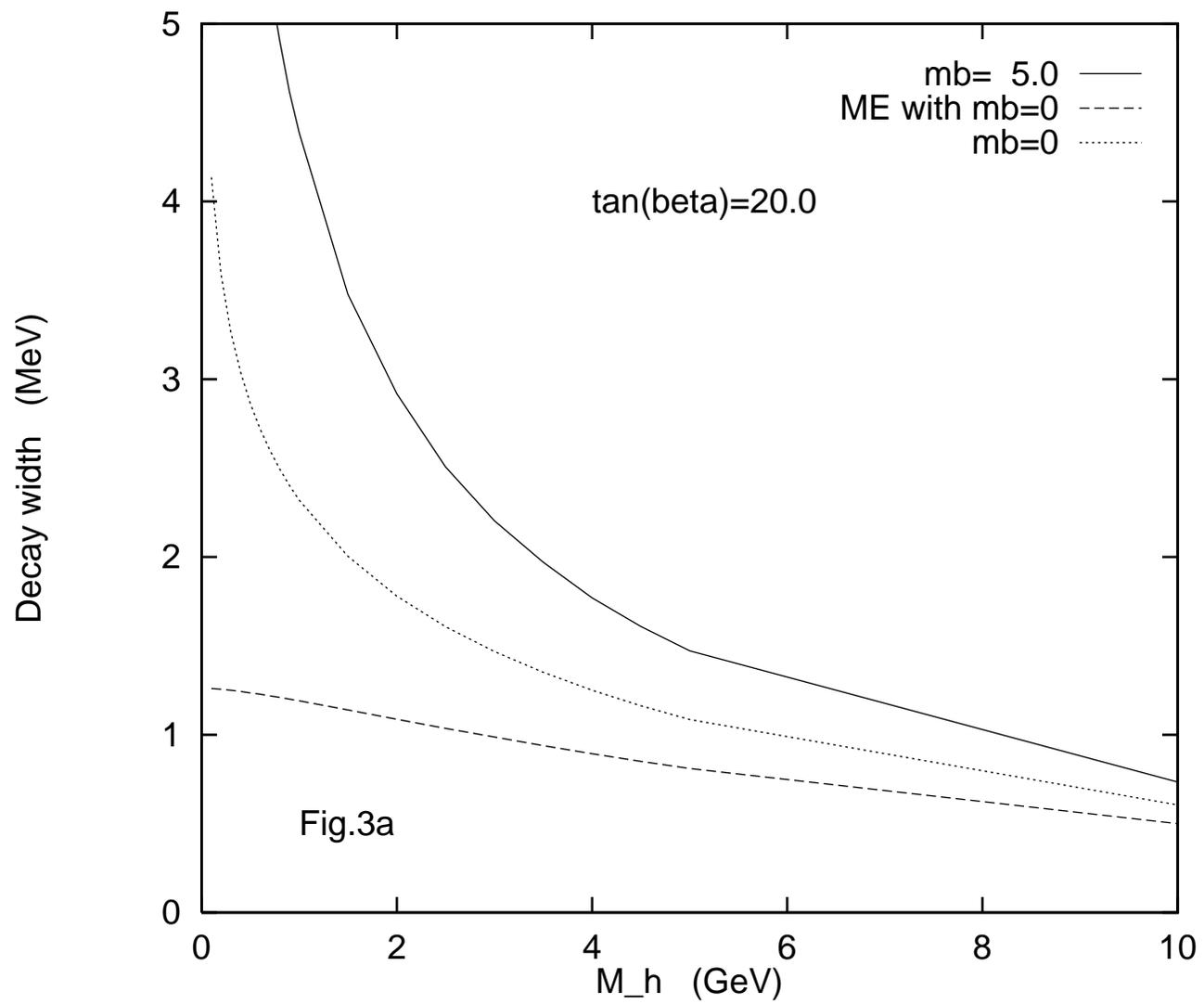
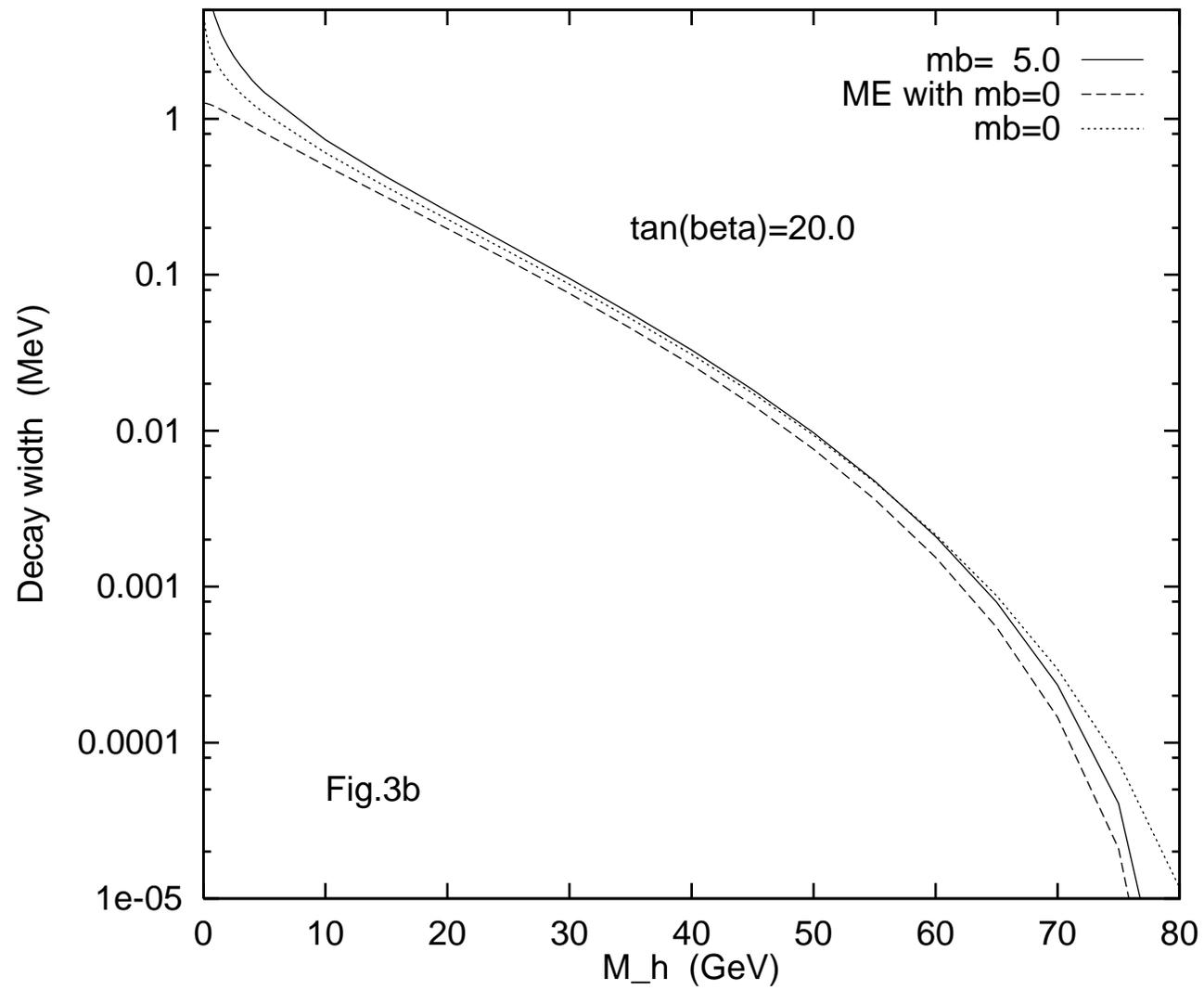
\end{document}